\documentstyle[12pt]{article}
\begin{document}
\thispagestyle{empty}
\begin{center}
\LARGE \tt \bf{Geodesics on non-Riemannian

 Geometric theory of Planar Defects}
\end{center}
\vspace{2.5cm}
\begin{center} {\large L.C. Garcia de Andrade\footnote{Departamento de
F\'{\i}sica Te\'{o}rica - Instituto de F\'{\i}sica - UERJ

Rua S\~{a}o Fco. Xavier 524, Rio de Janeiro, RJ

Maracan\~{a}, CEP:20550-003 , Brasil.

E-Mail.: GARCIA@SYMBCOMP.UERJ.BR}}
\end{center}
\vspace{2.0cm}
\begin{abstract}
The method of Hamilton-Jacobi is used to obtain geodesics around non-
Riemannian planar torsional defects.It is shown that by perturbation 
expansion in the Cartan torsion the geodesics obtained are parabolic curves 
along the plane x-z when the wall is located at the plane x-y.In the 
absence of defects the geodesics reduce to straight lines.The family of 
parabolas depend on the torsion parameter and describe a gravitationally 
repulsive domain wall.Torsion here plays the role of the Burgers vector in 
solid state physics.
\end{abstract}
\newpage
The investigation of topological defects \cite{1}in cosmology follows 
closed analogous models in solid state physics \cite{2}.Riemannian 
\cite{3,4,5} and non-Riemannian \cite{6,7} models have been used to 
investigate these distributional defects such as disclinations \cite{8}
and dislocations \cite{9} and domain walls analogous to ferromagnetic 
domains in crystallography.Lower dimensional models like disclinations in 
liquid crystals presented helical torsion structures as demonstrated by 
E.Dubois-Violette \cite{10} and James Sethna \cite{11}.More recently 
F.Moraes \cite{8} has been used the same method of Hamilton-Jacobi(H-J) 
equations to investigate geodesics around torsional defects like electrons 
around dislocated metals. In this Letter I follow this same patern to obtain 
geodesics around domain walls in 3-Dimensional gravity. Although isolated 
domain walls are nowadays of no interest in physics they still may play a 
role in mixed system in the condensed matter systems which serve as a 
laboratory for Cosmology.In this letter we find a simple example of a two 
dimensional gravity model from a domain wall solution of Einstein-Cartan 
gravity obtaing by reducing the dimensions of the spacetime metric 
\begin{equation}
ds^{2}= e^{8{\pi}G{\sigma}z}dt^{2}- e^{-{J}^{1}z}(dx^{2}+dy^{2})
-dz^{2}
\label{1}
\end{equation}
where $ J^{1} $ is the constant torsion at the planar wall,$ {\sigma} $
is the planar wall distribution of the wall energy density and G is the
Newtonian gravitational constant. Due to translational symmetry of 
defects we shall consider the following metric
\begin{equation}
ds^{2}= - e^{-{J}^{1}z}dx^{2}-dz^{2}
\label{2}
\end{equation}
at the $ y=constant $ section. To obtain the geodesics I shall solve 
the H-J equation
\begin{equation}
\frac{{\partial}W}{{\partial}t}+H(\frac{{\partial}W}{{\partial}x^{i}})=0
\label{3}
\end{equation}
where
\begin{equation}
W[{\gamma}]= \frac{1}{2} {\int}^{t_{1}}_{t_{0}}{g}_{ij} \dot{x}^{i} \dot{x}^{j}{dt}
%\label{4}
\end{equation}
where $ {\gamma} $ is a parametrized curve where
\begin{equation}
s[{\gamma}] = {\int}^{t_{1}}_{t_{0}} ({g}_{ij}{\dot{x}^{i} \dot{x}^{j})}^{\frac{1}{2}}dt
%\label{5}
\end{equation}
Here $ H = L = \frac{1}{2}{g_{ij} \dot{x}^{i} \dot{x}^{j}} $ is the 
Hamiltonian and $ L $ is the Lagrangian. We must stress that although in 
general Riemann-Cartan spaces test particles follow autoparallels and 
not geodesics ,off the domain wall there is vacuum and there GR remains
valid. Nevertheless geodesics suffer indirectly the influence of torsion
since they are also present in the metric considered. In the case of 
linearized EC gravity we have the following metric
\begin{equation}
ds^{2}=-({1-{J}^{1}z})dx^{2}-dz^{2}
\label{6}
\end{equation} 
The H-J equation applied to metric (\ref{6}) reduces to 
\begin{equation}
\frac{{\partial}W}{{\partial}t}+(1+J^{1}z)(\frac{{\partial}W}{{\partial}x})^{2}+(\frac{{\partial}W}{{\partial}z})^{2}=0
\label{7}
\end{equation}
where we have considered $ H $ in terms of the momenta $ p_{i} = \frac{{\partial}W}{{\partial}x^{i}} = \frac{{\partial}L}{{\partial}\dot{x}^{i}} $. Follows from (\ref{7}) that
\begin{equation}
W={c}_{1}t + W'(x,z) 
\label{8}
\end{equation}
and this in turn implies 
\begin{equation}
c_{1}+(\frac{{\partial}W'}{{\partial}x})^{2}+(\frac{{\partial}W'}{{\partial}z})^
{2}=0
\label{9}
\end{equation}
Writing the perturbation series for $ W'$ as  
\begin{equation}
W'={W'}_{0}+J^{1}{W'}_{1}+(J^{1})^{2}{W'}_{2}+... 
\label{10}
\end{equation}
and substituting (\ref{10}) into (\ref{9}) one obtains
\begin{equation}
({\frac{{\partial}{W'}_{0}}{{\partial}x}})^{2}+({\frac{{\partial}{W'}_{0}}{{\partial}z}})^{2}=-c_{1}
\label{11}
\end{equation}
and
\begin{equation}
-\frac{{\partial}{W'}_{0}}{{\partial}x}^{2}z+2\frac{{\partial}{W'}_{1}}{{\partial}z}\frac{{\partial}{W'}_{0}}{{\partial}z}+2\frac{{\partial}{W'}_{1}}{{\partial}x}\frac{{\partial}{W'}_{0}}{{\partial}x}=0
\label{12}
\end{equation}
Equation (\ref{11}) is solved at once by the ansatz
\begin{equation}  
{W'}_{0}={c}_{2}x+{c}_{3}z 
\label{13}
\end{equation}
where the constants satisfy 
\begin{equation}
{c}_{1}=-\frac{1}{2}({c_{2}}^{2}+{c_{3}}^{2})
\label{14}
\end{equation}
Now the equation (\ref{12}) can be written as
\begin{equation}
c_{2}\frac{{\partial}{W'}_{1}}{{\partial}x}+{c}_{3}\frac{{\partial}{W'}_{1}}{{\partial}y}=-
\frac{1}{2}{c_{2}}^{2}z 
\label{15}
\end{equation}
Equation (\ref{15}) is solved by the method of carachteristics where we 
first differentiate $ {W'}_{1} $ with respect to the auxialiry variable 
$ u $ such as 
\begin{equation}
\frac{d{W'}_{1}}{du}=\frac{{\partial}x}{{\partial}u}\frac{{\partial}{W'}_{1}}{{\partial}x}+\frac{{\partial}{W'}_{1}}{{\partial}z}\frac{{\partial}z}{{\partial}u}
\label{16}
\end{equation}
Comparison of (\ref{15}) with (\ref{16}) yields
\begin{equation}
\frac{d{W'}_{1}}{du}=\frac{1}{2}{c_{2}}^{2}z
\label{17}
\end{equation}
and
\begin{equation}
\frac{{\partial}x}{{\partial}u}=c_{2}
\label{18}
\end{equation}
and
\begin{equation}
\frac{{\partial}z}{{\partial}u}=c_{3}
\label{19}
\end{equation}
Thus 
\begin{equation}
x=c_{2}u+t_{1}
\label{20}
\end{equation}
and
\begin{equation}
z=c_{3}u
\label{21}
\end{equation}
where $t_{1}$ is another parameter. Now by integration of the equation 
(\ref{17})we obtain
\begin{equation}
{W'}_{1}=-\frac{1}{2}{c_{3}}^{4}u^{2}+d
\label{22}
\end{equation}
Where $ d $ is an integration constant. Thus by substituting (\ref{22}) 
into (\ref{10}) one has
\begin{equation}
W=-\frac{1}{2}({c_{2}}^{2}+{c_{3}}^{2})t+c_{2}x+c_{3}z+J^{1}d+\frac{1}{2}{c_{3}}^{4}
J^{1}u^{2}
\label{23}
\end{equation}
The solution of the motion equations is obtained by differentiation 
with respect to $ c_{2} $ and $ c_{3} $  of $ W $. Thus
\begin{equation}
0=\frac{{\partial}W}{{\partial}c_{3}}=-{c}_{3}t + z-4{c_{3}}z^{2}J^{1}
\label{24}
\end{equation}
and
\begin{equation}
0=\frac{{\partial}W}{{\partial}{c}_{2}}=-{c}_{2}t+ x
\label{25}
\end{equation}
Equations (\ref{24}) and (\ref{25}) yield the following geodesic equations
\begin{equation}
x=c_{2}t
\label{26}
\end{equation}
and
\begin{equation} 
z=-\frac{c_{3}x}{c_{2}}+4{J^{1 }}{c_{3}}z^{2}
\label{27}
\end{equation}
which yields the desired trajectory. From (\ref{27}) one observes 
that torsion  contributes to the parabolic curve.In the absence of the 
torsional defect the geodesics are reduced to the straight lines.Our 
result describes a gravitationally repulsive domain wall is algebraically simpler than the one obtained by F.Moraes due to the simpler geometry of the torsional defect we are dealing with. Future prospects in this work include the investigation of autoparallels in propagating torsion  theories.After I finished this paper I came across with an interesting paper by Katanaev and Volovich \cite{12} where they considered elastic waves on parallel wedge dislocations in the geometric theory of defects also dealing with the Hamilton-Jacobi methods.Nevertheless they did not handle domain walls as we did here.
\section*{Acknowledgement}
I would like to express my gratitude to Prof.H.Kleinert for kindly 
suggesting this work to me and for his kind hospitality at the 
Institut fur theoretische physik at Berlin Freie Universitat. Thanks 
are also due Prof. P.S.Letelier for helpful discussions on the subject 
of this Letter.Special thanks go to Prof.M.O.Katanaev and Prof.James Sethna for sending me their papers.Financial support from CNPq.(Brazil) and DAAD (Bonn) is 
grateful ackowledged.


\newpage
\begin{thebibliography}{16}
\bibitem{1} A. Vilenkin and P.S. Shellard, Cosmic String and Other 
Topological Defects, (1993), Cambridge Univ. Press. 
\bibitem{2} E.Kroner,Continuum theory of Defects,course 3,in Physics of 
Defects,Les Houches Session XXXV,North-Holland,Amsterdam,(1981).
\bibitem{3} P.S.Letelier,J.Math.Phys.(1995).
\bibitem{4} P.S.Letelier, Class. and Quant. Grav., (1995),\underline{12},2221.
\bibitem{5} L.C.Garcia de Andrade,J.Math.Phys.(1998),Jan.issue.
\bibitem{6} L.C.Garcia de Andrade,Mod.Phys.Lett.A,12,27(1997).
\bibitem{7} L.C.Garcia de Andrade,On non-Riemannian domain walls,Gen. Rel. and Grav.(1998),in press.
\bibitem{8} F.Moraes,Phys. Lett. A 214,(1996)189.
\bibitem{9} L.C.Garcia de Andrade,Non-Riemannian cosmic walls as boundaries of spinning polarized matter,Mod.Phys.Lett.A,in press.
\bibitem{10}E.Dubois-Violette and B.Pansu,Geometry of Condensed Matter Physics,Directions in condensed matter physics vol.9,(1990),World Scientific. 
\bibitem{11}J.P.Sethna,Phys.Rev.B31,(1985),10,6278.
\bibitem{12}M.O.Katanaev and I.V.Volovich,Scattering on dislocations and cosmic string in the geometric theory of defects,Los Alamos preprint archives gr-qc/9801081.
\end{thebibliography}
\end{document}